\documentclass[%
 aip,
 amsmath,amssymb,
 reprint,%
]{revtex4-1}

\usepackage{graphicx}
\usepackage{dcolumn}
\usepackage{bm}

\usepackage[utf8]{inputenc}
\usepackage[T1]{fontenc}
\usepackage{mathptmx}

\usepackage[mathscr]{eucal} 
\usepackage{xcolor}

\begin{document}

\preprint{AIP/123-QED}

\title{Hetero-structure Mode Space Method for Efficient Device Simulations}

\author{Mincheol Shin}
\email{mshin@kaist.ac.kr}
\affiliation{ 
School of Electrical Engineering,
Korea Advanced Institute of Science and Technology\\
291 Daehak-ro, Yuseong-gu, Daejeon 34141, Rep. of Korea 
}%


\date{\today}

\begin{abstract}
The Hamiltonian size reduction method or the mode space method applicable to general heterogeneous structures is developed in this work. The effectiveness and accuracy of the method are demonstrated for four example devices of GaSb/InAs tunnel field effect transistor (FET), MoTe$_2$/SnS$_2$ bilayer vertical FET, InAs nanowire FET with a defect, and Si nanowire FET with rough surfaces. The Hamiltonian size is reduced to around 5 \% of the original full Hamiltonian size without losing the accuracy of the calculated transmission and local density of states in a practical sense. The method developed in this work can be used with any type of Hamiltonian and can be applied to virtually any hetero-structure, so it has the potential to become an enabling technology for efficient simulations of hetero-structures.
\end{abstract}

\maketitle

%

\section{\label{sec:introduction}Introduction\protect}


As transistors continuously shrink and novel devices adopting new materials and innovative operation principles constantly emerge, there is a growing need for accurate modeling and simulation of hetero-structure devices. III-V hetero-junction field effect transistors (FETs),\cite{Riel2014} two-dimensional (2D) bilayer vertical FETs,\cite{Roy2014} metal/ferroelectric/metal junctions\cite{Tsymbal2006} and nanowire interconnects/contact\cite{Wang2008} are a few example hetero-structures among many. For their atomistically-resolved modeling, the first-principles density function theory (DFT) method has clear advantages over the empirical models such as the tight-binding (TB) model or the k$\cdot$p theory. In the parameter-free DFT method, the ambiguous, inaccurate, and often impossible parameter fitting processes are avoidable.


DFT Hamiltonians have found increasing use in the nanoelectronics simulations recently. Si nanowire FETs\cite{Shin2016,Bruck2017} and ultra-thin-body FETs,\cite{Bruck2017} 2D material FETs,\cite{Szabo2015,Ahn2019,Kim2020,Afzalian2021} and resistive memory devices\cite{Ducry2020} were simulated by importing LCAO DFT Hamiltonians, where LCAO is an abbreviation for linear combinations of atomic orbitals. Si nanowire transistors were also simulated by using the Hamiltonian from the real-space DFT method.\cite{Mori2020} As for the plane wave basis DFT method, 2D material channel FETs were simulated through constructing the mode space Hamiltonian.\cite{Pala2020} As a practical approach suited for treating realistically sized devices, these studies make use of the Hamiltonian imported from an equilibrium DFT calculation, separated from the non-equilibrium Green's function (NEGF) transport calculation. On the other hand, more rigorous DFT approach that employs the non-equilibrium DFT and NEGF methods\cite{Taylor2001} is practically limited to molecule sized devices, although there has been a recent progress with the scattering state calculation approach.\cite{Ye2021}


A disadvantage of adopting DFT Hamiltonian in device simulations is the computational burden of handling large-sized DFT Hamiltonian whose interaction range is often longer than simple nearest neighbor interaction. Thus it is desirable to come up with an effective Hamiltonian constructed with much smaller set of basis than the original full set of basis. The Hamiltonian size reduction method or more widely known as the mode space (MS) method is attractive in that the computational time and resources can be reduced by a few orders of magnitude, while the errors in the charge density and current are kept within a few percents or less.

The MS method has been developed for a wide range of Hamiltonians: effective mass theory,\cite{Venugopal2002,Shin2007} k$\cdot$p method,\cite{Shin2009} TB,\cite{Milnikov2012} pseudo-potential,\cite{VandePut2019,Pala2019} and DFT Hamiltonians.\cite{Shin2016,Pala2020} The MS method essentially relies on the Bloch states of periodic structures so it has been mostly applied to homogeneous structures where there is one kind of unit cell. It has also been applied to simple hetero-structures such as the A-B type hetero-structure of Fig.\ \ref{fig:hetero_type} (b) (but without the junction cells JA and JB), but the single-unit-cell MS method was just applied to each of materials A and B separately.\cite{Chen2020} To the best of our knowledge, there has been no attempt to treat the hetero-structures with junction cells as shown in Fig.\ \ref{fig:hetero_type} (b) or the hetero-structures with irregular regions as shown in Fig.\ \ref{fig:hetero_type} (c). (Refer to Section \ref{sec:hetero_types} for detailed description of the hetero-structures of Fig.\ \ref{fig:hetero_type}.)

Unlike the unit cells of homogeneous materials, junction cells or cells in the irregular region are not meant to be repeated periodically, so the very concept of the electronic band associated with the cells is not pertinent nor the basis matrix constructed with the Bloch wave functions. That means that the single-unit-cell MS method cannot be applied to the individual cells. The salient feature of the hetero-structure MS method developed in this work is that it does not demand nor exploit the periodicity of the individual cells of a hetero-structure so individual cells can assume any arbitrary form. An extreme example of this is Si nanowire with rough surfaces in the channel region as shown in Fig.\ \ref{fig:devices_examples} (d) and Fig.\ \ref{fig:Si_SR}.


In this work, we introduce a MS scheme for hetero-structure or supercell consisting of heterogeneous cells, extending our previous work of the MS method for single unit cell.\cite{Shin2016} Although we demonstrate the method with LCAO DFT and TB Hamiltonians in this work, the method is not limited to particular types of Hamiltonian. Therefore the MS method developed in this work should be also applicable to DFT Hamiltonian in the plane-wave or real-space basis, k$\cdot$p Hamiltonian and other model Hamiltonians.

\section{\label{sec:hetero_types}Types of Hetero-structure\protect}

\begin{figure}[tb]
\includegraphics[width=\columnwidth]{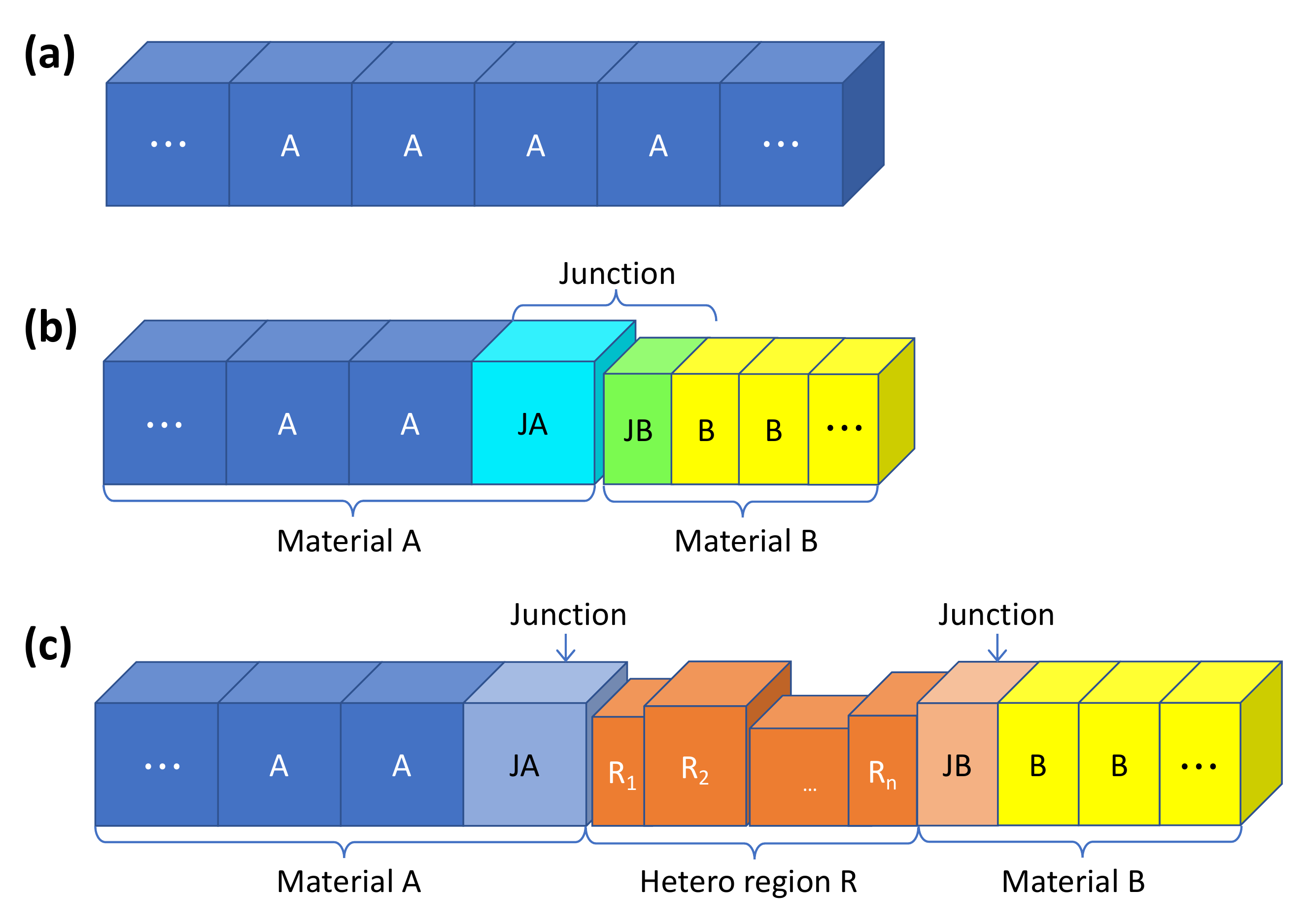}
\caption{\label{fig:hetero_type} (a) Homogeneous structure with unit cell A. (b) The A-B type hetero-structure: unit cells A and B are the ones for homogeneous materials A and B, respectively, and JA and JB are the junction cells at the contact. (c) The A-R-B type hetero-structure: in addition to the unit cells and junction cells of (b), there exists a hetero region between two materials, represented by cells $\rm R_1$, $\rm R_2$, .., $\rm R_n$.}
\end{figure}

\begin{figure}[tb]
\includegraphics[width=\columnwidth]{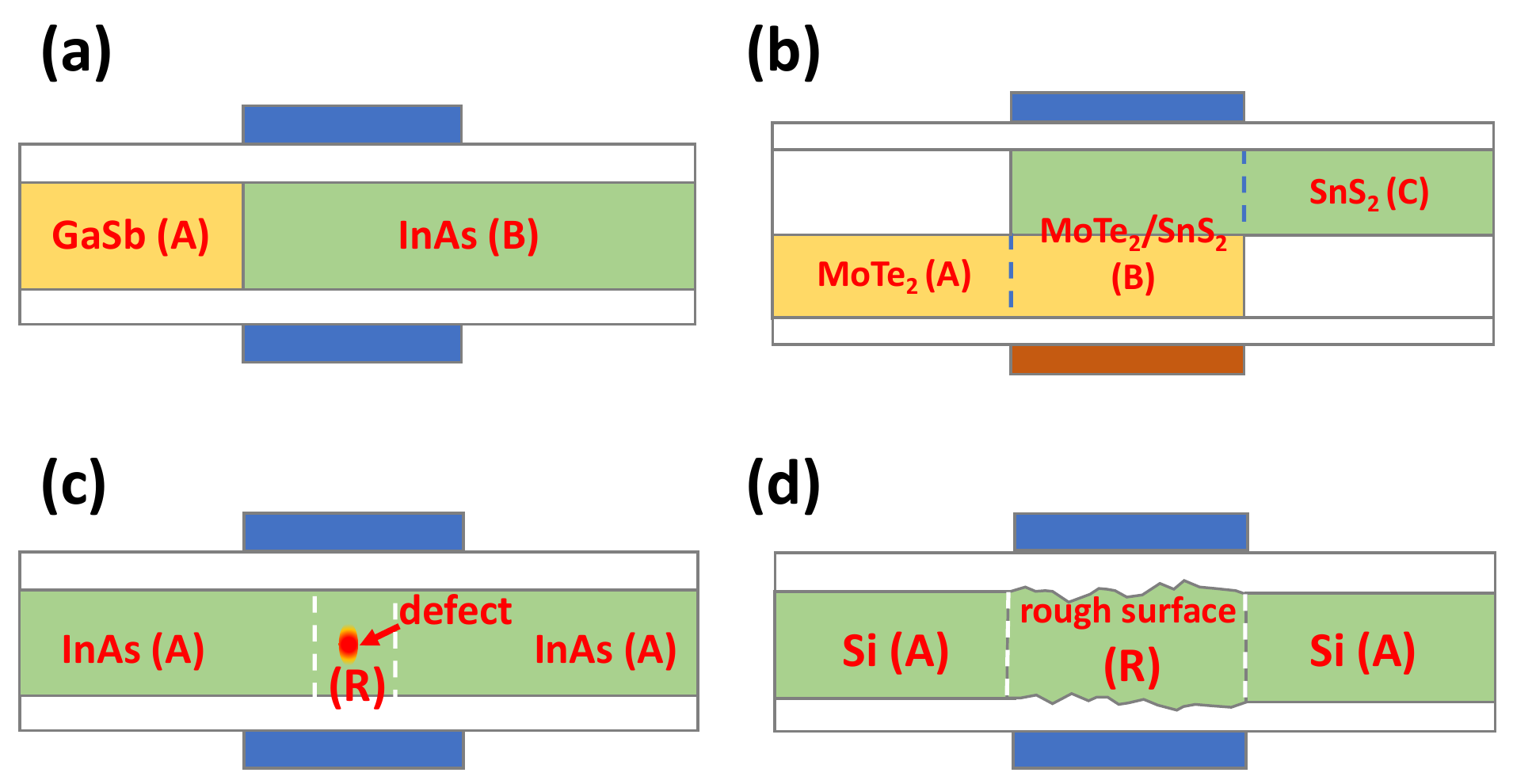}
\caption{\label{fig:devices_examples} (a) GaSb/InAs hetero-junction tunnel FET as an example of the A-B type hetero-structure. (b) MoTe$_2$/SnS$_2$ bilayer vertical tunnel FET as an example of the A-B-C type hetero-structure. (c) InAs nanowire FET with an As antisite defect as an example of the A-R-B hetero-structure (B=A). (d) Si nannowire FET with rough surfaces in the channel region as an extreme example of the A-R-B hetero-structure (B=A).}
\end{figure}

In a hetero-structure, at least two kinds of unit cells are needed. Figs.\ \ref{fig:hetero_type} (b) and (c) show two types of hetero-structures considered in this work: the A-B type and the A-R-B type. Firstly, in the A-B type, there are two kinds of unit cells A and B for the respective materials, and junction cells JA and JB at the junction. Junction cell JA is regarded a spatially transient cell from the unit cell A, having the same number of atoms and orbitals as the unit cell A but the atom positions are different. It also serves as a buffer cell which blocks the influence of material B on material A so that the unit cell property of cell A is kept intact. The similar applies to junction cell JB. The GaSb/InAs hetero-structure shown in Fig.\ \ref{fig:devices_examples} (a) and Fig.\ \ref{fig:GaSb_InAs} is an example system of the A-B hetero-structure.

As a straightforward extension of the A-B hetero-structure, the A-B-C structure is also possible to be simulated, where there exist a third material with unit cell C and also junction cells between B and C. The hetero-structure MoTe2/SnS2 shown in Fig.\ \ref{fig:devices_examples} (b) and Fig.\ \ref{fig:MoTe2_SnS2} is an A-B-C type hetero-structure.

The second hetero-structure type considered in this work is the A-R-B type as shown in Fig.\ \ref{fig:hetero_type} (c). It consists of two materials with their respective unit cells A and B and a hetero region R in between them. The cells in the hetero region (marked with R$_1$, R$_2$, ... R$\rm{_n}$ in the figure) can be different from cells A and B in terms of the atom species, the number of atoms, and atom positions and etc, so the hetero region may contain defects or consist of atom-thin layers of different kinds or have irregular shaped cells. It is also possible that there exist junction cells JA and JB between A and R and between R and B, respectively. Examples of the A-R-B type are InAs nanowire with a point defect shown in Fig.\ \ref{fig:devices_examples} (c) and Fig.\ \ref{fig:InAs_trap} and Si nanowire with surface roughness in the channel region shown in Fig.\ \ref{fig:devices_examples} (d) and Fig.\ \ref{fig:Si_SR}. In fact, the A-B type hetero-structure can be regarded as a subset of the A-R-B type hetero-structure, but it is convenient to distinguish into the two types as the A-B type structures are quite common in practice.

\section{\label{sec:homo_MS}Brief Summary on Single-cell MST\protect}

Hereafter, the MS method for TB Hamiltonian\cite{Milnikov2012} and its extension to DFT Hamiltonian\cite{Shin2016} is called `single-cell mode space technique (MST).' The single-cell MST is briefly summarized in this Section.

A homogeneous structure can be constructed by indefinitely repeating a unit cell as shown in Fig.\ \ref{fig:hetero_type} (a) and its electronic bands naturally arise from exploiting the periodicity. In the single-cell MST, a set of the Bloch wave functions are selected from the electronic bands within a preset energy window of interest and used to construct the basis matrix. The basis matrix contains, by its construction, all the physical states or the Bloch states that were used to form the basis matrix. For the cases of TB and DFT Hamiltonians, the basis matrix usually generates `unphysical' or spurious states as well. The art of single-cell MST lies in removing the unphysical states systematically through a minimization method. After all the unphysical states are removed, the basis matrix retains only the physical states with which the electronic bands are faithfully reproduced in the preset energy window. 

For the minimization procedure, a variational functional $F$ is introduced which is to be minimized upon adding an extra basis $U'$ to the basis matrix. Through the relationship of $U'=\Xi C$ where $\Xi$ is the matrix called the trial function and $C$ is a column vector, the problem is reduced to finding the vector $C$ which minimizes $F$. Usually one unphysical branch is removed by one extra basis, so the procedure is repeated until all the unphyiscal states are removed. 

In short, the single-cell MST has two key ingredients: 1) formation of the initial basis matrix with the physical states within an energy window of interest, and 2) minimization of the variational functional. The method relies on the periodicity of the unit cell so it can be applied only to homogeneous structures.

\section{\label{sec:approach}Hetero-structure MST\protect}

In this Section, the detailed method to reduce the Hamiltonian size of hetero-structures is described. The hetero-structure mode-space technique is in principle the same as the single-cell MST but there are some significant modifications in its implementation, which are (1) the MS method applied to a supercell, and (2) construction of the basis matrix in a block diagonal form.

\subsection{\label{sec:supercell}Supercell}

The first step in our hetero-structure simulation is to devise a `supercell' whose individual cells are used to construct the desired hetero-structure and its Hamiltonian. This step is especially important for the DFT method as the supercell Hamiltonian is constructed concurrently. For the TB or k$\cdot$p method, the supercell is not needed for the Hamiltonian construction itself but it is still needed for the hetero-structure MST.

\begin{figure}[tb]
\includegraphics[width=\columnwidth]{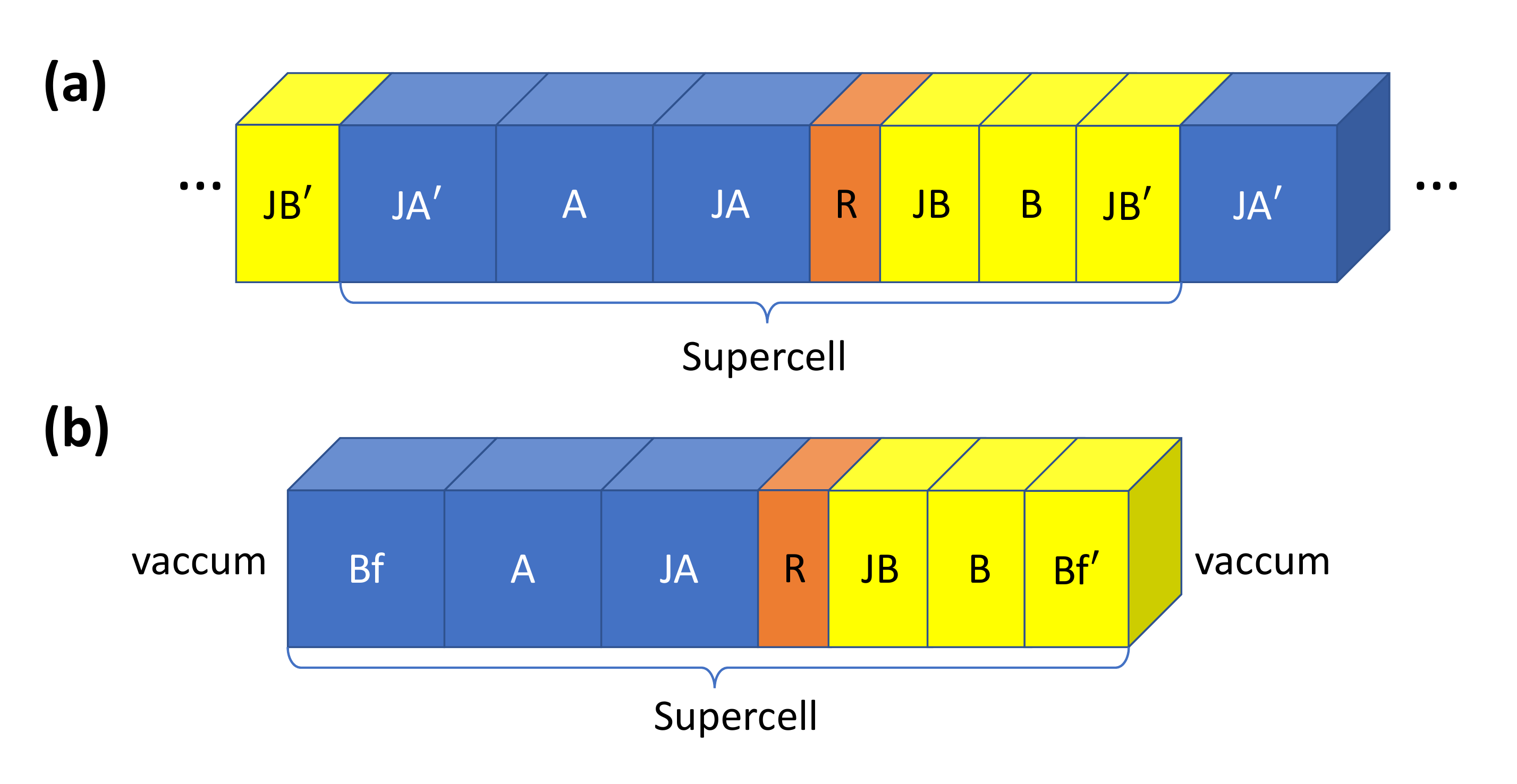}
\caption{\label{fig:supercell_kind} (a) Supercell with the periodic boundary condition. (b) Molecule-like supercell with vacuum regions placed at both ends. In (a) and (b), cells A and B are regarded {\it unit cells} of homogeneous materials.}
\end{figure}

Suppose that we simulate a A-R-B type hetero-structure shown in Fig.\ \ref{fig:hetero_type} (c) with one cell in the hetero region R. A supercell for it should include cells A, JA, R, JB, and B which are arranged contiguously as they are in the hetero-structure device. For the DFT Hamiltonian construction, we either impose periodic boundary conditions on the supercell or place vacuum regions to both ends of the supercell. For the former case, B-A junction is needed in addition so that the supercell would consist of JA$'$, A, JA, R, JB, B, JB$'$ cells as shown in Fig.\ \ref{fig:supercell_kind} (a), where junction cells JB$'$ and JA$'$ are the ones for B-A junction. For the latter case, buffer cells Bf and Bf$'$ need to be placed at both ends so that the supercell would consist of Bf, A, JA, R, JB, B, Bf$'$ cells as shown in Fig.\ \ref{fig:supercell_kind} (b). As aforementioned, the junction or buffer cells show spatially transient behaviors and also serve as blocking cells, which means that there should be sufficient number of junction or buffer cells in order for the cells A and B to be qualified as the unit cells of homogeneous materials in the practical sense. In other words, if the cell A is isolated from the supercell and repeated indefinitely to make a homogeneous structure, the resulting electronic band structure should match with that of the intended homogeneous material. The same applies to cell B. Among the cells in a supercell, the cells which possess the homogeneous-material unit-cell quality are called {\it unit cells} and others just cells hereafter. We comment that a unit cell of a homogeneous material is usually synonymous with unit cell which is pristine (without any defects) and has ideal surfaces (without dangling bonds). 

As a concrete example of a supercell and {\it unit cells} in it, see GaSb/InAs hetero-structure and the supercell for it in Fig.\ \ref{fig:GaSb_InAs}. There are a total of 10 cells in the periodic supercell: two {\it unit cells} A (GaSb) and B (InAs) and 4 junction cells for each junction. By repeating {\it unit cells} A and B and placing junction cells between them, the GaSb/InAs hetero-structure device with the source, channel and drain regions can be realized. Another example is MoTe$_2$/SnS$_2$ vertical tunnel FET shown in Fig.\ \ref{fig:MoTe2_SnS2}. The supercell for it has 15 cells; three {\it unit cells} A (MoTe$_2$), B (MoTe$_2$/SnS$_2$) and C (SnS$_2$), which respectively are used to make up the source, channel, and drain regions, and 4 junction cells at each of A-B junction and B-C junction, and two buffer cells at each of the two ends that are in contact with vacuum regions.

\subsection{\label{sec:objective}Objective}

Suppose that a supercell prepared as described in the previous subsection consists of $L$ cells and that the supercell Hamiltonian is written as:
\begin{equation}
    \mathscr{H}^{sc}=\left[
    \begin{array}{ccccc}
         H_{11}& H_{12} & H_{13} & ... &H_{1,L}  \\
         H_{21} & H_{22} & H_{23} & ... & H_{2,L} \\
         ... & ... & ... & ... & .. \\  
        H_{L,1} & H_{L,2} & H_{L,3} & ... & H_{L,L}
    \end{array}
    \right]
    \label{eq:supercell_Hamiltonian}
\end{equation}
where the block element $H_{b,b'}$ is the inter-cell Hamiltonian between cells $b$ and $b'$ or intra-cell Hamiltonian if $b=b'$, where $1\le b, b' \le L$. $H_{b,b'}$ has the size $N_b\times N_{b'}$ where $N_b$ is the Hamiltonian size of cell $b$ and the total size of the supercell Hamiltonian is $N = \sum_b N_b$.

The objective of the hetero-structure MST is to unitarily transform the supercell Hamiltonian by a block diagonal basis matrix $\mathscr{U}$ such that the transformed Hamiltonian $h^{sc}$ given by
\begin{equation}
    h^{sc}=\mathscr{U}^\dagger \mathscr{H}^{sc} \mathscr{U},
\end{equation}
has the same entries as the original supercell Hamiltonian but with reduced size. Specifically, suppose that the supercell basis matrix $\mathscr{U}$ is of block diagonal form given by
\begin{equation}
    \mathscr{U}=\left[
        \begin{array}{ccccc}
        U_1&0&0&0&0\\
        0&U_2&0&0&0\\
        0&0&...&0&0\\
        0&0&0&...&0\\
        0&0&0&0&U_L
        \end{array}
    \right]
    \label{eq:U_blockdiagonal}
\end{equation}
where its block diagonal element $U_b$ has the size $N_b \times n_b$ where $n_b$ is the number of modes of cell $b$ and $n_b < N_b$ $(1 \le b \le L)$. Then
\begin{equation}
    h^{sc}=\left[
    \begin{array}{ccccc}
         h_{11} & h_{12} & h_{13} & ... & h_{1,L}  \\
         h_{21} & h_{22} & h_{23} & ... & h_{2,L} \\
         ... & ... & ... & ... & .. \\  
         h_{L,1} & h_{L,2} & h_{L,3} & ... & h_{L,L}
    \end{array}
    \right]
\end{equation}
where
\begin{equation}
    h_{b,b'}=U_b^\dagger H_{b,b'} U_{b'}.
\end{equation}

The hetero-structure MST applied to a supercell, or the supercell MST, is essentially the same as the single-cell MST. The difference lies in forming the initial basis matrix in a block diagonal form and designing the trial function such that the basis matrix $\mathscr{U}$ retains the block diagonal form.

\subsection{\label{sec:initial_basis}Initial Basis Matrix}

In this subsection, we describe how to set up the supercell initial basis matrix in a block diagonal form of Eq.\ (\ref{eq:U_blockdiagonal}). We begin by setting up the supercell initial basis matrix in a non-block-diagonal form.

If the supercell is a non-periodic molecule structure, we solve for the eigenstates of the supercell given by 
\begin{equation}
\mathscr{H}^{sc} \Psi_{v} = E_{v}\mathscr{S}^{sc} \Psi_{v}
\label{eq:supercell_molecule}
\end{equation}
where $\mathscr{H}^{sc}$ is the supercell Hamiltonian matrix of Eq.\ (\ref{eq:supercell_Hamiltonian}) and $\mathscr{S}^{sc}$ is the corresponding supercell overlap matrix with block elements $S_{b,b'}$ defined similarly to $H_{b,b'}$. For orthogonal system, $S_{b,b'}=I \delta_{b,b'}$ where $I$ is the identity matrix. If the supercell is a periodic structure where cell 1 comes to the right of cell $L$ periodically, we solve the supercell band-structure equation which is given by (for a nanowire structure)
\begin{equation}
\mathscr{H}_k \Psi_{v,k} = E_{v, k}\mathscr{S}_k \Psi_{v,k}   
\label{eq:supercell_band}
\end{equation}
where 
\begin{eqnarray}
    \mathscr{H}_k &=& \mathscr{H}_0 + e^{ika_{sc}} \mathscr{W}+e^{-ika_{sc}} \mathscr{W}^{\dagger}, \nonumber \\
    \mathscr{S}_k &=& \mathscr{S}_0 + e^{ika_{sc}} \mathscr{S}_1+e^{-ika_{sc}} \mathscr{S}_1^{\dagger}, 
\end{eqnarray}
where $\mathscr{H}_0 = \mathscr{H}^{sc}$, $\mathscr{S}_0 = \mathscr{S}^{sc}$ and $\mathscr{W}$ and $\mathscr{S}_1$ are the Hamiltonian and overlap matrices that couple adjacent supercells and $a_{sc}$ is the spatial length of the supercell. To be specific, for $L = 5$ and with the interaction limited to nearest cells, 
\begin{equation}
    \mathscr{H}_0 = \left[
    \begin{array}{ccccc}
    H_{11}&H_{12}&0&0&0\\
    H_{21}&H_{22}&H_{23}&0&0\\
    0&H_{32}&H_{33}&H_{34}&0\\
    0&0&H_{43}&H_{44}&H_{45}\\
    0&0&0&H_{54}&H_{55}
    \end{array}\right],
\end{equation}

\begin{equation}
    \mathscr{W} = \left[
    \begin{array}{ccccc}
    0&0&0&0&0\\
    0&0&0&0&0\\
    0&0&0&0&0\\
    0&0&0&0&0\\
    H_{51}&0&0&0&0
    \end{array}
    \right],
\end{equation}
and $\mathscr{S}_0$ and $\mathscr{S}_1$ are similarly given. In this work, the nearest-cell interaction is assumed for all the four example devices selected for demonstration, but the hetero-structure MST described in this work is not limited to the nearest-cell interaction.

We solve Eq.\ (\ref{eq:supercell_molecule}) or (\ref{eq:supercell_band}) and select a set of eigenstates or Bloch states $\{\Psi_\mu\}$ within an energy window of interest, where $\mu = 1,..,m$ where $m$ is the number of selected eigenstates or Bloch states, and then construct a unitary matrix which has $\Psi_\mu$'s as its column vectors, 
\begin{equation}
    \mathscr{U}^{\rm ND} = \left[\Psi_1, \Psi_2, ...,\Psi_m\right],
    \label{eq:Usc}
\end{equation}
which is an initial basis matrix of the supercell in the non-block-diagonal (ND) form. $\mathscr{U}^{\rm ND}$ has size $N\times m$. These steps are identical to the steps for setting up the initial basis matrix by the single-cell MST,\cite{Shin2016} except that a larger-sized supercell is dealt with in place of a single unit cell. 
 
We divide the matrix $\mathscr{U}^{\rm ND}$ row-wise by blocks into $L$ sub-matrices, $U_1, U_2, .., U_L$, where $U_b$ has size $N_b \times m$ $(1\le b \le L)$. Recall that $N_b$ is the Hamiltonian size of cell $b$ and $N=\sum_b N_b$. We then form a block diagonal matrix $\mathscr{U}$ by placing the sub-matrices $U_b$'s at the diagonal entries. Lastly, $U_b$'s are ortho-normalized to have the size $N_b \times n_b$ where $n_b < m$. As a concrete example, for the $L=5$ case, 
\begin{equation}
    \mathscr{U}^{\rm ND}=\left(
    \begin{array}{cc}
         U_1  \\
         U_2  \\
         U_3  \\
         U_4 \\
         U_5
    \end{array}
    \right)
    \Rightarrow
    \mathscr{U}=\left[
        \begin{array}{ccccc}
        U_1&0&0&0&0\\
        0&U_2&0&0&0\\
        0&0&U_3&0&0\\
        0&0&0&U_4&0\\
        0&0&0&0&U_5
        \end{array}
    \right]
    \label{eq:U0_construction}
\end{equation}

We make the ansatz that $\mathscr{U}$ is equivalent to $\mathscr{U}^{\rm ND}$ in terms of their ability to contain all the physical states which were initially selected. In the single-cell MST the initial basis matrix reproduces the physical states by its construction, but it is not obvious how to form the initial basis matrix for a supercell with the condition of the block diagonal form imposed. What it means by the above mentioned ansatz is that $\mathscr{U}$ constructed as shown in Eq.\ (\ref{eq:U0_construction}) contains all the eigenstates or Bloch states that were initially chosen. Thus we can use it as an initial basis matrix to which we add extra modes to remove the spurious states. After all the spurious states are removed, the basis matrix will reproduce the energy spectrum or the electronic band structure within the preset energy window of interest, which is the goal of the MS method. 

\begin{figure}[tb]
\includegraphics[width=\columnwidth]{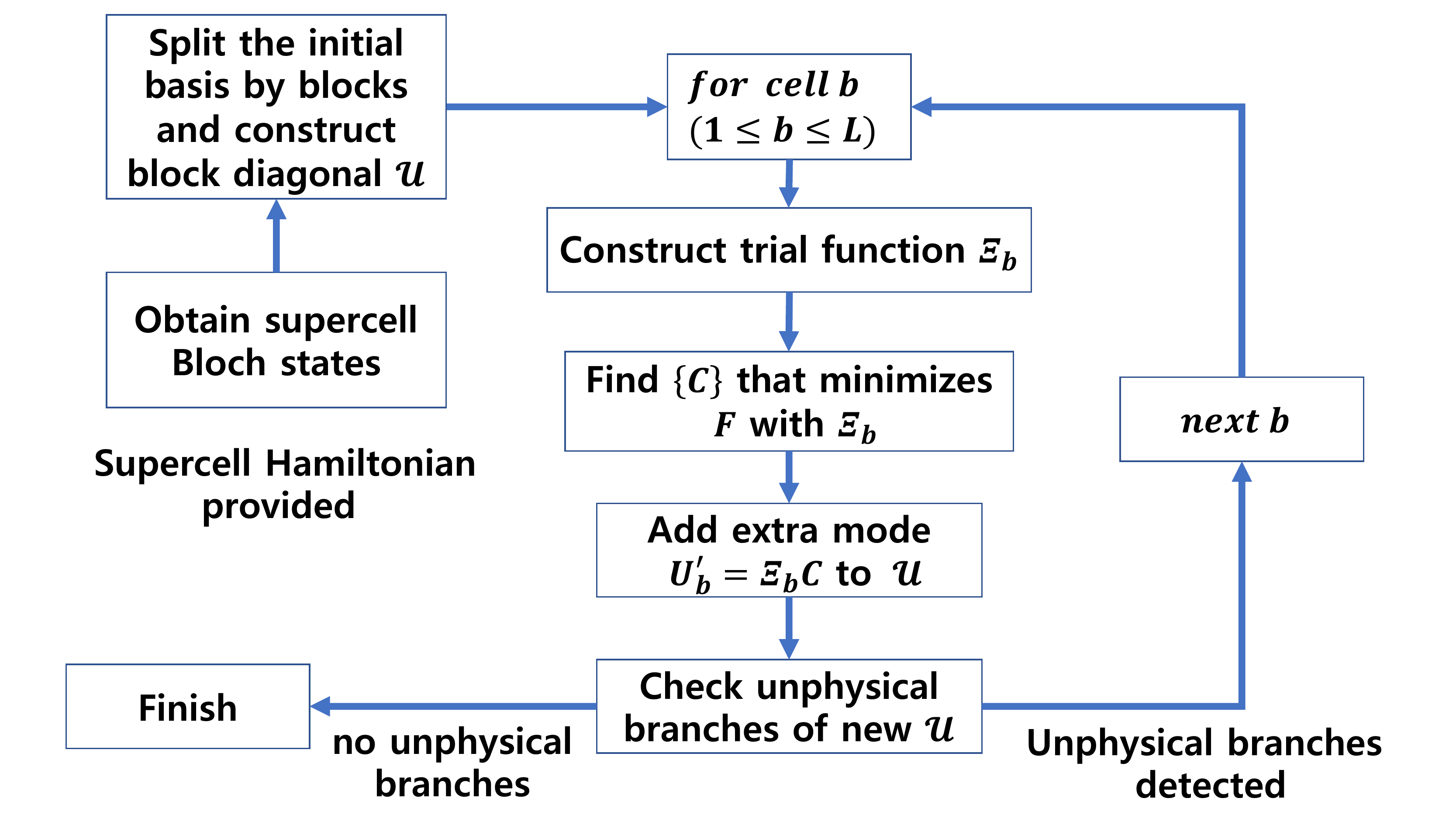}
\caption{\label{fig:flowchart} Flow chart for the hetero-structure MST.}
\end{figure}

\subsection{\label{sec:trial_function}Trial Function}

In the hetero-structure MST, the trial function should be designed in such a way that the basis matrix $\mathscr{U}$ should maintain the block tridiagonal form during the iterations for unphysical state removal. For the purpose, while there should be other ways to design the trial functions, we have chosen the trial function in a simple yet effective way as follows. 

For a target cell $b$ ($1\le b\le L$) we construct the trial function $\Xi_b$ which is a matrix with size $N\times N_{\xi,b}$ where $N_{\xi,b}$ is to be determined later. The $b$-th block row of $\Xi_b$ is a submatrix $\xi_b$ of size $N_b \times N_{\xi,b}$ which is given by
\begin{equation}
    \xi_b=u_b [\mathscr{S}_{k=0}^{-1}\mathscr{H}_{k=0}, \mathscr{S}_{k=\pi}^{-1}\mathscr{H}_{k=\pi} ]\mathscr{U},
        \label{eq:xi_b}
\end{equation}
and all the rest block rows are zero. In the above equation, the matrix $u_b$ of size $N_b \times N$ has its $b$-th block column given by $(1-U_b U_b^\dagger)$ and all the rest block columns are zero. In Eq.\ (\ref{eq:xi_b}), $A[B,C]D$ is a short-hand expression of $[ABD, ACD]$ where $A, B, C, D$ are matrices and $[A,B]$ means concatenation of two matrices $A$ and $B$. After ortho-normalization, $\xi_b$ becomes of the size $N_b\times N_{\xi,b}$ where $N_{\xi,b} \le 2n$ where $n=\sum_b n_b$ is the column size of the basis matrix $\mathscr{U}$.

As a concrete example, for the 5-cell supercell and $b=3$,
\begin{equation}
    \Xi_3 = \left[
    \begin{array}{c}
    0\\
    0\\
    u_3 [\mathscr{S}_{k=0}^{-1}\mathscr{H}_{k=0}, \mathscr{S}_{k=\pi}^{-1}\mathscr{H}_{k=\pi} ]\mathscr{U}\\
    0\\
    0\\
    \end{array}
    \right]
\end{equation}
and 
\begin{equation}
    u_3=\left[ 
    \begin{array}{ccccc}
    0 & 0 & (1-U_3 U_3^\dagger) & 0 & 0
    \end{array}
    \right]
\end{equation}

This form for the trial function is a straightforward extension of the trial function for the single-cell MST\cite{Shin2016} and in particular it ensures that the supercell transformation matrix $\mathscr{U}$ maintains the diagonal form of Eq.\ (\ref{eq:U_blockdiagonal}). With $\Xi_b$ as the trial function, the minimization function $F$ is constructed exactly the same way as in the case of the single-cell MST and the parameter set $\{C\}$ that minimizes $F$ is obtained.\cite{Milnikov2012,Shin2016} Then the column vector for an extra mode $U_b'=\Xi_b C$ is calculated and added to $\mathscr{U}$. Thanks to how $\Xi_b$ is constructed, $U_b'$ is added only to the transformation matrix for cell $b$ and thus the diagonal form of $\mathscr{U}$ is maintained. The rest of the steps are the same as the single-cell case. The target cell $b$ is changed sequentially (one cell after another) or in some designated order to facilitate the removal process of the unphysical states. Once all the unphysical states are removed, we get the transformation matrix for each cell $U_b$. See Fig.\ \ref{fig:flowchart} for the overall flow. 

As a technical note, among the cells belonging to the supercell, any cell $\tilde{b}$ which is designated as a {\it unit cell} receives special treatment as follows. Recall how the term {\it unit cell} is used in this paper by referring to Section \ref{sec:supercell}. The cell $\tilde{b}$ is isolated from the supercell and the single-cell MST is applied to the cell to obtain the final basis matrix (the basis matrix after all the unphysical states are cleared). In the initial basis matrix $\mathscr{U}$ of the supercell, the diagonal block $(\tilde{b},\tilde{b})$ is replaced with the final basis matrix from the single-cell MST. And, during the subsequent iterations, the extra modes are not added to the cell $\tilde{b}$ so that the basis matrix corresponding to the cell is left intact. In other words, for the cells designated as the {\it unit cells}, we use the basis matrices determined by the single-cell MST, and the basis matrix for other cells (junction cells, buffer cells, or cells in the hetero region) are determined by the supercell MST. This has been proven to be more efficient and reliable.

\section{\label{sec:applications}Applications of the hetero-structure MST\protect}

In this Section, we apply the hetero-structure MST to four different hetero-structure devices and demonstrate the effectiveness and accuracy of the method. We used our in-house ballistic NEGF transport simulator to calculate the transmission and local density of states (LDOS). The detailed methods employed in our simulator to calculate the quantities can be found in Ref. [\onlinecite{Shin2016}]. The quantities calculated by using the full real-space Hamiltonian and the ones calculated by using the reduced, effective Hamiltonian by the hetero-structure MST are compared.

\subsection{\label{sec:GaSb_InAs}GaSb/InAs hetero-junction tunnel FET}

\begin{figure*}[tb]
\includegraphics[width=\textwidth]{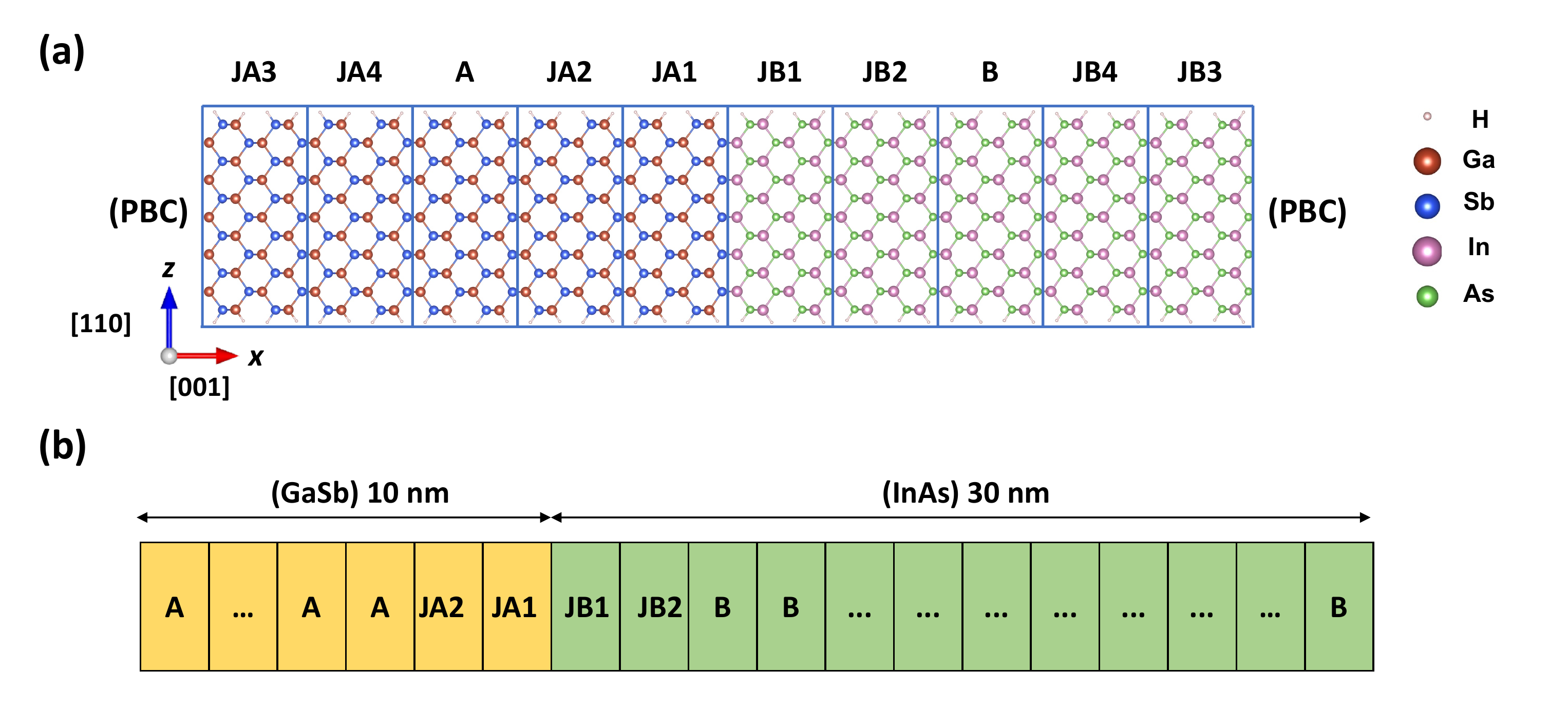}
\caption{\label{fig:GaSb_InAs} GaSb/InAs hetero-junction tunnel FET: (a) The 10-cell supercell used in the hetero-structure MST. The periodic boundary conditions (PBC) are imposed to the supercell. JA3, JA4, A, JA2, and JA1 are GaSb cells while the rest InAs cells. Cells A and B are {\it unit cells} of GaSb and InAs, respectively, and other cells are junction cells. (b) The device region for the simulation of the GaSb/InAs tunnel FET of Fig.\ \ref{fig:devices_examples} (a) with 10 nm GaSb source, 10 nm InAs channel, and 20 nm InAs drain regions.}
\end{figure*}

\begin{figure*}[tb]
\includegraphics[width=\textwidth]{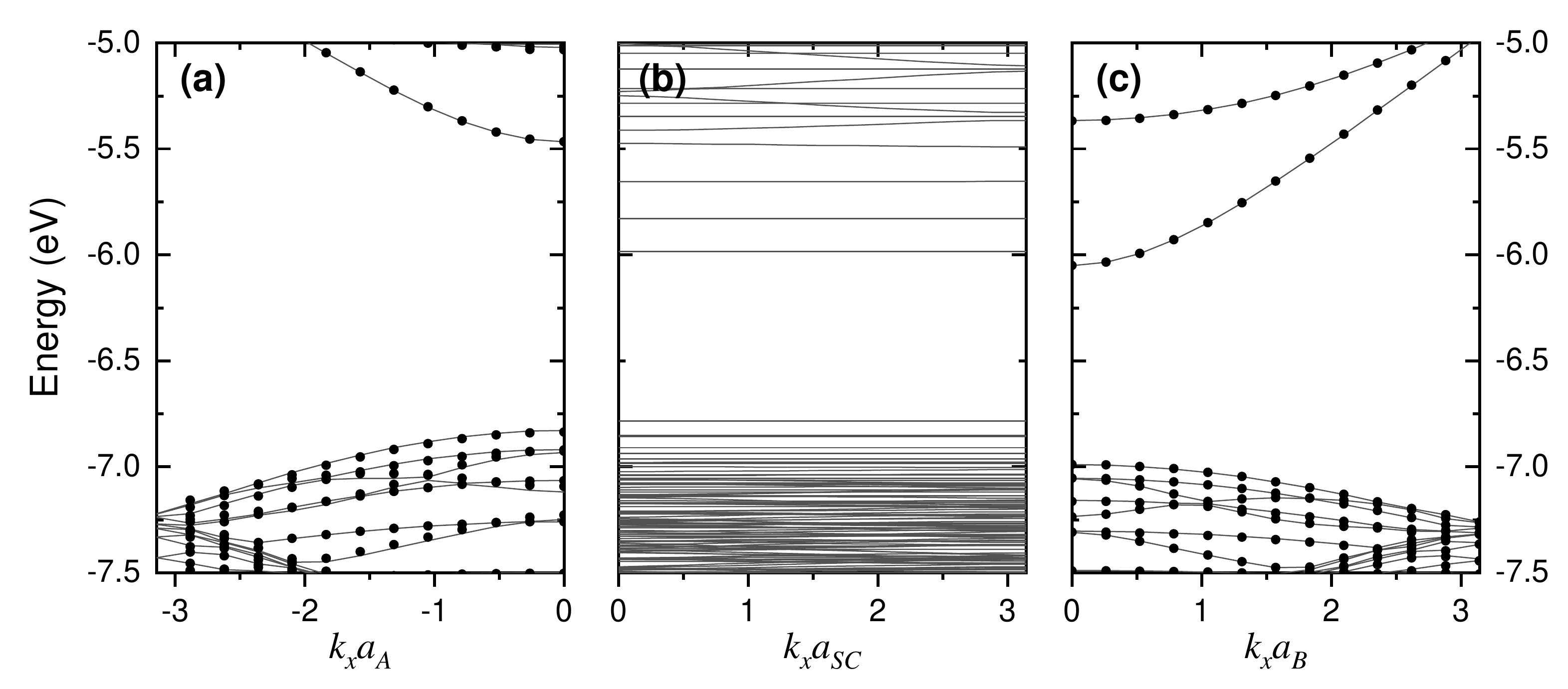}
\caption{\label{fig:GaSb_InAs_band} (a) Thin lines represent the electronic bands of the cell A isolated from the supercell of Fig.\ \ref{fig:GaSb_InAs} (a). Solid dots are for the bands of unstrained homogeneous GaSb of 2.4 nm thickness. (b) The band structure of the 10-cell supercell of Fig.\ \ref{fig:GaSb_InAs} (a). (c) Thin lines are for the bands of the cell B isolated from the supercell. Solid dots are for the bands of the unstrained homogeneous InAs of 2.4 nm thickness. $a_A$, $a_B$, and $a_{SC}$ are the length of the cells A and B and the supercell, respectively. The longitudinal wave vector $k_y = 0$.}
\end{figure*}

\begin{figure*}[tb]
\includegraphics[width=\textwidth]{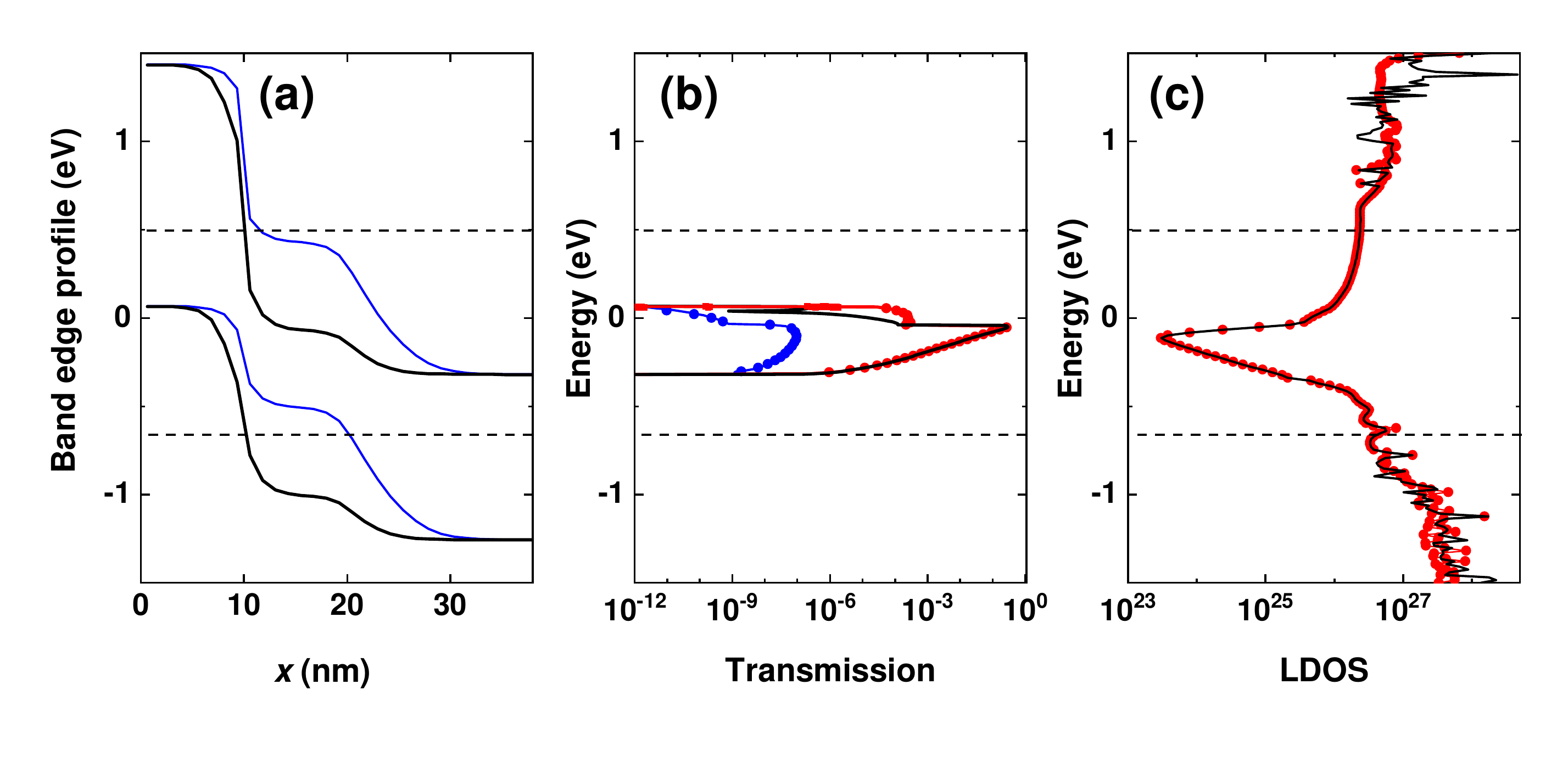}
\caption{\label{fig:GaSb_InAs_transmission} (a) The band-edge profiles of the 2.4 nm double-gate UTB GaSb/InAs tunnel FET at the two gate voltages of $V_g = 0\,V$ (thin blue lines) and $0.5\,V$ (thick black lines). The longitudinal wave vector $k_y$ is set to 0. (b) The transmission calculated by using the MS Hamiltonian (dotted lines) and by using the original Hamiltonian (solid lines) at $V_g = 0\,V$ (blue) and $0.5\,V$ (black for the original and red for the MS), (c) LDOS calculated by using the MS Hamiltonian (red dotted line) and by using the original Hamiltonian (black solid line) at the hetero-junction located at $x = 10$ nm (cell JB1) at $V_g = 0.5\,V$. The energy window of interest used for the hetero-structure MST is indicated by dashed lines.}
\end{figure*}

\begin{table}
\caption{\label{table:GaSb_InAs} The full-Hamiltonian size $N_b$ and the number of modes $n_b$ of the cells of the supercell of Fig.\ \ref{fig:GaSb_InAs} (a). A total of 13 unphysical branches were cleared during the hetero-structure MST procedure.}
\begin{ruledtabular}
\begin{tabular}{ccccccc}
cell & A & JA2 & JA1 & JB1 & JB2 & B \\
\hline
$N_b$ & 832 & 832 & 832 & 832 & 832 & 832 \\
$n_b$ & 60 & 60 & 41 & 42 & 46 & 46
\end{tabular}
\end{ruledtabular}
\end{table}

As an example of the A-B type hetero-structure, we simulated a double-gate ultra-thin-body (UTB) GaSb/InAs tunnel FET of Fig.\ \ref{fig:devices_examples} (a), with the channel thickness of 2.4 nm, channel length of 10 nm, and 10 nm and 20 nm source and drain regions, respectively. The GaSb/InAs junction is located at the source/channel junction. The device region for the simulation is shown in Fig.\ \ref{fig:GaSb_InAs} (b) where there are 8 GaSb cells and 24 InAs cells (8 cells for the channel and 16 cells for the drain). 

The supercell for the hetero-structure MST is constructed as shown in Fig.\ \ref{fig:GaSb_InAs} (a), where there are a total of 10 cells with 5 cells for GaSb and InAs each. As we simulated an UTB structure, the structure is repeated in the y direction and its periodicity is taken care of by introduction of the longitudinal wave vector $k_y$ in the Brillouin zone. In the z-direction, the surface atoms are passivated with hydrogen atoms. 

The 10-cell supercell is constructed in such a way that the cells marked with A and B should be qualified as {\it unit cells}. That is, cells A and B should be close to the unit cells of homogeneous material A (GaSb) and homogeneous material B (InAs), respectively, in terms of their atom positions and Hamiltonian entries, and consequently the electronic band structure. To achieve that there are four junction cells between A and B cells which constitute the GaSb/InAs junction. As the supercell is constructed to be periodic in the x direction, there is also the B-A junction with four junction cells. The size of each cell is twice that of a primitive unit cell of homogeneous material, containing 22 Ga(In) atoms, 22 Sb(As) atoms, and 8 hydrogen atoms in each cell, because the interaction range between atoms is greater than 5th neighbors and the supercell Hamiltonian is constructed to be of block tri-diagonal form. 

The 10-cell supercell was relaxed by the DFT method using the SIESTA package.\cite{Soler2002} PSML norm-conserving pseudopotentials\cite{Garcia2018} and the generalized gradient approximation (GGA) and Perdew, Burke and Ernzerhof exchange (PBE) functional\cite{Perdew1996} were employed. The band gap underestimation of GGA-PBE were adjusted by using the DFT-1/2 technique.\cite{Ferreira2008} The bulk and slab structures used to construct the supercell were fully relaxed until the maximum force became less than 0.05 eV/\r{A}.

After relaxation, we verified that the cells A and B are indeed qualified as {\it unit cells} by drawing the band structures calculated by using the cells isolated from the supercell. See Figs.\ \ref{fig:GaSb_InAs_band} (a) and (c), respectively, where the band structure of the cell A (B) is drawn and compared with the band structure of the unit cell of {\it unstrained} homogeneous GaSb (InAs) of the same thickness. The band structures for InAs match perfectly, while there are slight differences in the band structures for GaSb due to the slight mismatch of the lattice constants at the interface of the two materials: GaSb in the supercell was compressed by 0.8\% in order for its lattice constant to match with that of InAs at the interface. Otherwise, the match would have been perfect for GaSb as well.

The hetero-structure MST was carried out on the supercell, with the energy windows of VBE-0.7 and CBE+0.7 for the supercell, which corresponds to the energy window of VBE-0.7 and CBE+0.2 for cell A and the energy window of VBE-0.5 and CBE+0.8 for cell B, where VBE$-\alpha$ and CBE$+\beta$ means $\alpha$ eV below the valence band edge (VBE) and $\beta$ eV above the conduction band edge (CBE). The resulting number of modes are shown in Table \ref{table:GaSb_InAs}, where the average number of modes per cell is 49, which is about 6\% of the size of full Hamiltonian. 
 
We have tested the accuracy of the hetero-structure MST by comparing the current density and LDOS. Fig.\ \ref{fig:GaSb_InAs_transmission} shows the band-edge profiles of the UTB GaSb/InAs tunnel FET for two gate voltages, and corresponding energy-resolved current density and LDOS. LDOS are shown for the junction cell JB1. The current density and LDOS calculated by using the MS Hamiltonian through the hetero-structure MST and the ones by using the full Hamiltonian match with each other excellently.

\begin{figure*}
\includegraphics[width=\textwidth]{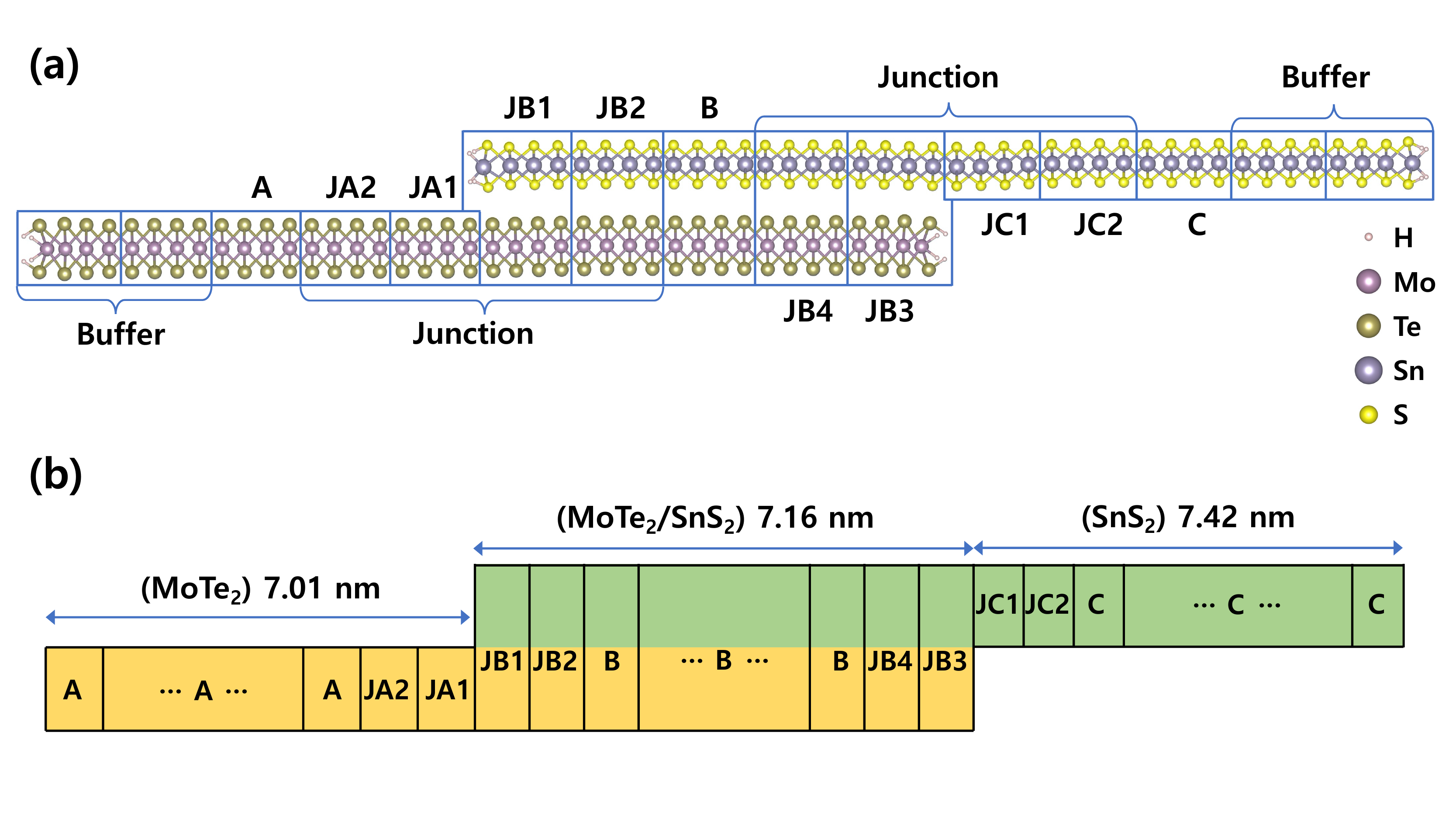}
\caption{\label{fig:MoTe2_SnS2} MoTe$_2$/SnS$_2$ bilayer vertical tunnel FET: (a) The 15-cell supercell used in the hetero-structure MST. The bottom and top layers are MoTe$_2$ and SnS$_2$, respectively. Both ends of the supercell were hydrogen passivated as well as the dangling bonds in the junction cells JB1 and JB3. (b) The device region for the simulation of the MoTe$_2$/SnS$_2$ vertical tunnel FET of Fig.\ \ref{fig:devices_examples} (b).
}
\end{figure*}

\begin{figure*}
\includegraphics[width=\textwidth]{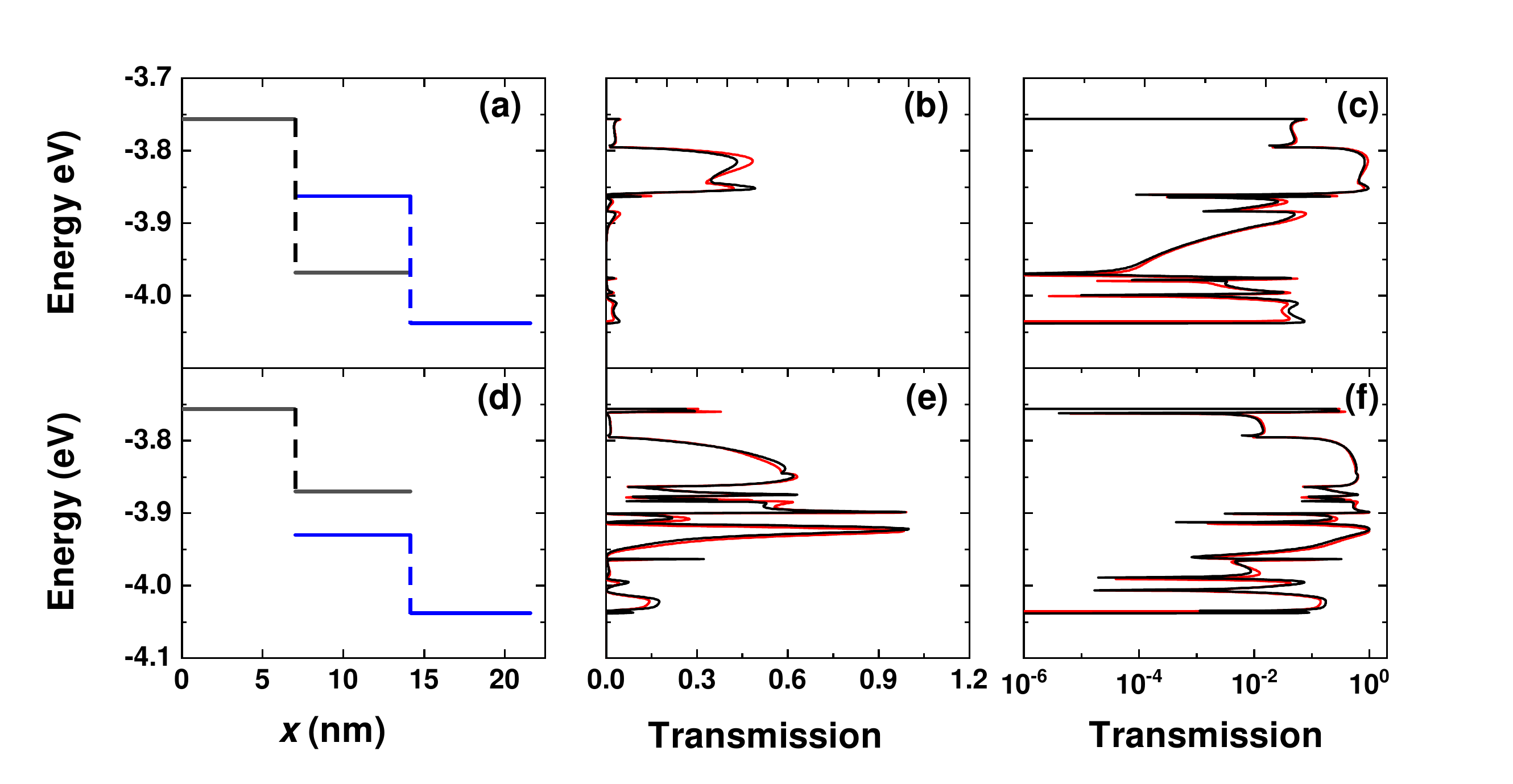}
\caption{\label{fig:MoTe2_SnS2_TR} (a), (b), and (c) are for OFF state and (d), (e), and (f) are for ON state. (a) and (d): the band-edge profile of the MoTe$_2$/SnS$_2$ system. Black lines are for the valence band edge of MoTe$_2$ and blue lines the conduction band edge of SnS$_2$. (b) and (e): the transmission in the linear scale. (c) and (f): the transmission in the log scale. In (b), (c), (e) and (f), black lines represent the transmission calculated by using the full Hamiltonian and red lines the one by using the MS Hamiltonian. The longitudinal wave vector $k_y$ is set to zero.}
\end{figure*}

\begin{figure*}
\includegraphics[width=\textwidth]{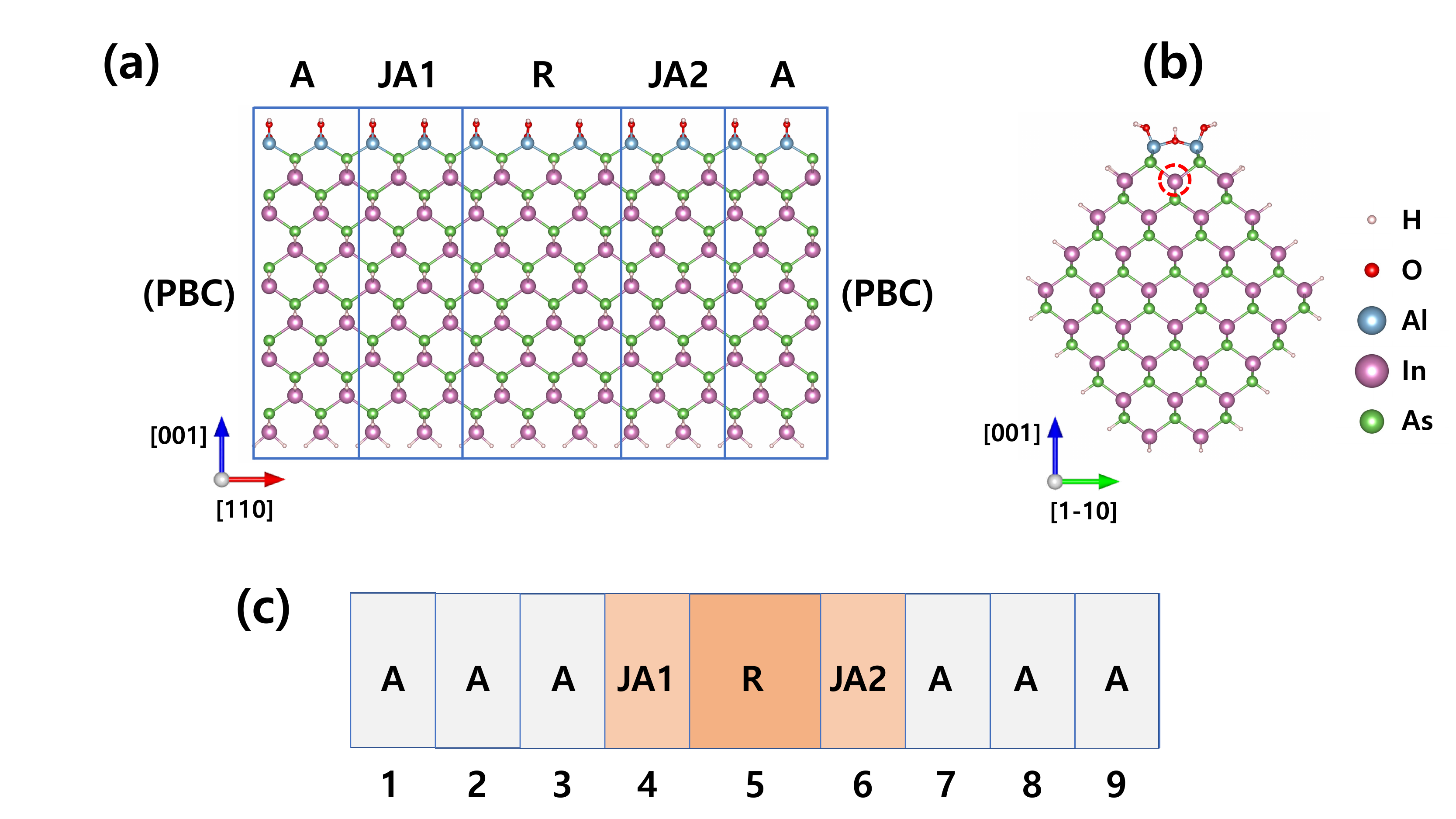}
\caption{\label{fig:InAs_trap} InAs nanowire FET with an As antisite defect: (a) The 5-cell supercell used in the hetero-structure MST. The periodic boundary condition is imposed to the supercell. (b) Cross-sectional view of the supercell. The atom position where the As anti-site defect is located is indicated by dashed circle. (c) The device structure used for the transmission calculation. The 9 cells are numbered as displayed.}
\end{figure*}

\begin{figure*}
\includegraphics[width=\textwidth]{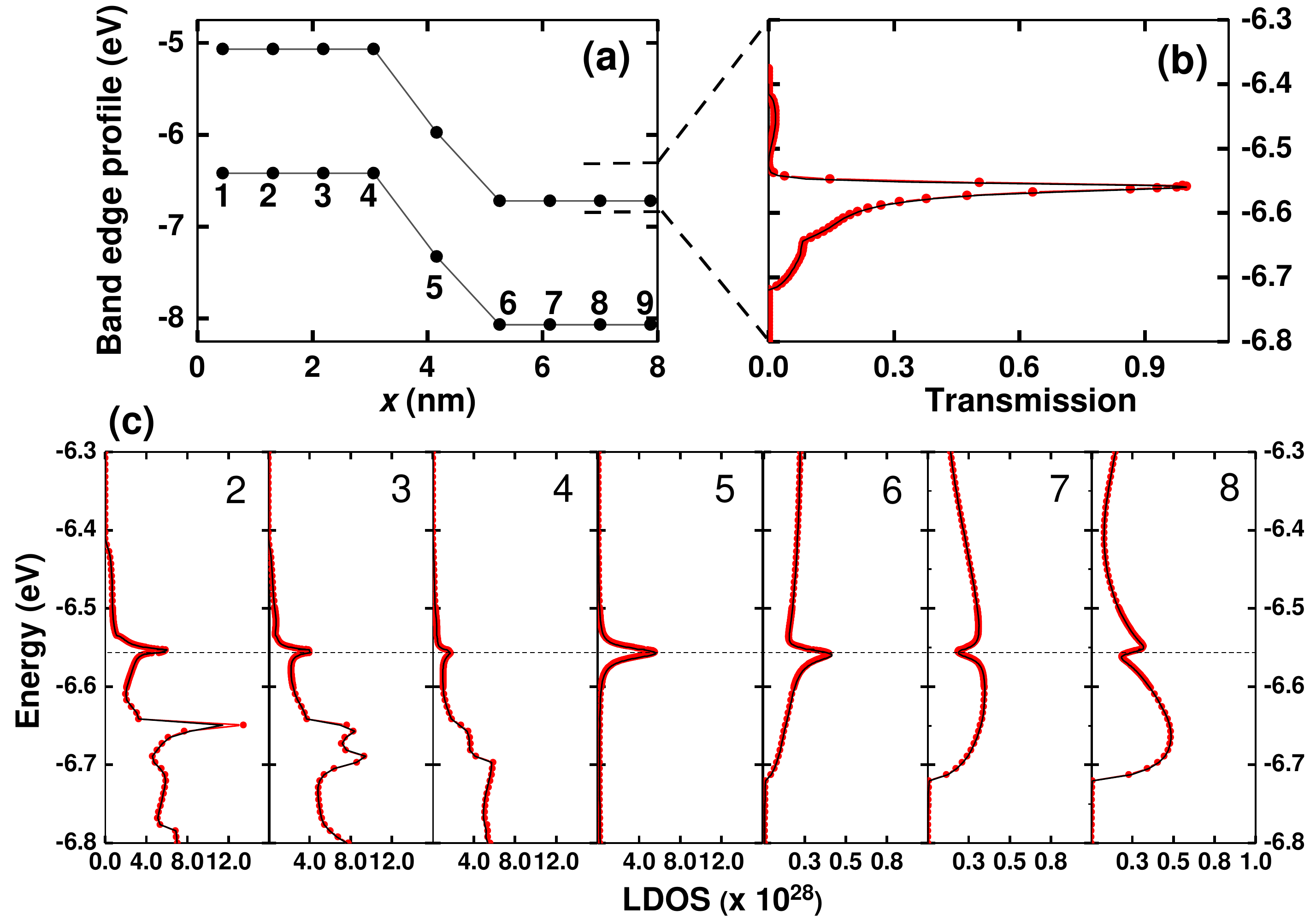}
\caption{\label{fig:InAs_trap_TR} (a) The band edge profile used for the transmission and LDOS calculations of the device structure of Fig.\ \ref{fig:InAs_trap} (c). (b) The transmission and (c) LDOS calculated by using the MS Hamiltonian (red dots) and by using the full Hamiltonian (black lines). In (c), the cell number is indicated at the top right corner of each panel. The dashed line in (c) indicates the location of the trap state due to the As antisite defect.}
\end{figure*}

\subsection{\label{sec:MoTe2SnS2}MoTe$_2$/SnS$_2$ vertical tunnel FET}

\begin{table}
\caption{\label{table:MoTe2_SnS2}
The full-Hamiltonian size $N_b$ and the number of modes $n_b$ of the cells of the supercell of Fig.\  \ref{fig:MoTe2_SnS2} (a). A total of 39 unphysical branches were cleared during the hetero-structure MST procedure.}
\begin{ruledtabular}
\begin{tabular}{cccccccccccc}
cell & A & JA2 & JA1 & JB1 & JB2 & B & JB4 & JB3 & JC1 & JC2 & C \\
\hline
$N_b$ & 196 & 196 & 196 & 372 & 352 & 352 & 352 & 372 & 156 & 156 & 156 \\
$n_b$ & 47 & 47 & 43 & 50 & 51 & 59 & 50 & 39 & 17 & 11	& 20
\end{tabular}
\end{ruledtabular}
\end{table}

As an example system of the A-B-C type hetero-structure, we simulated MoTe$_2$/SnS$_2$ vertical tunnel FET shown in Fig.\ \ref{fig:devices_examples} (b). The supercell to which the hetero-structure MST is applied is shown in Fig. \ref{fig:MoTe2_SnS2} (a) where there are 15 cells, of which five left cells are monolayer MoTe$_2$, five center cells bilayer MoTe$_2$/SnS$_2$ and five right cells monolayer SnS$_2$. The cells marked with A, B, and C are regarded (and verified to be) {\it unit cells} and there are four junction cells at each of the left and right junctions. The left and right ends of the supercell are terminated with hydrogen atoms, making the supercell like a molecule structure. The two leftmost and two rightmost cells serve as buffer cells that block the effect of the termination.

There are 12 atoms in each of MoTe$_2$ and SnS$_2$ cells, and 24 for MoTe$_2$/SnS$_2$ bilayer. With the interaction range between atoms taken into account, the cell size is larger than that of smallest unit cell for each material in order to ensure that the supercell Hamiltonian has block tri-diagonal form. The supercell was relaxed by DFT using the SIESTA tool in a similar way to the one in Section \ref{sec:GaSb_InAs}. 

With the DFT Hamiltonian of the relaxed supercell, the supercell MST was carried out on the 15-cell supercell. Since the supercell is not periodic in the lateral direction, the eigenvalue problem in Eq. (\ref{eq:supercell_molecule}) is solved for the molecular-like structure instead of the usual band-structure calculation of Eq.\ (\ref{eq:supercell_band}). For the energy window of VBE-1.0 to CBE+1.4 for the supercell MST, the MS transformation matrix for each cell was successfully determined with the number of modes shown in Table \ref{table:MoTe2_SnS2}. 

Fig. \ref{fig:MoTe2_SnS2} (b) shows the structure for the transmission calculation, where the cells A, B, and C extracted from the supercell are repeated to construct the source, channel, and drain regions, respectively, each of which consists of 10 cells. We calculated the transmission for the two relevant situations in FET operations. The first one corresponds to OFF state where the conduction band (CB) of MoTe$_2$/SnS$_2$ is above the valence band (VB) so tunneling does not take place in the bilayer. Figs.\ \ref{fig:MoTe2_SnS2_TR} (b) and (c) show the transmission both in the linear and log scales. Note that the MS solutions through the hetero-structure MST match very well with the full solutions. The tunneling current is effectively suppressed in the band gap region of the bilayer between -3.97 and -3.86 eV. The pronounced transmission in the energy range of -3.86 and -3.8 eV leads to leakage current which should be suppressed using a space layer in practical applications.

The second situation corresponds to ON state where the inversion between CB and VB of MoTe$_2$/SnS$_2$ takes place as shown in Fig.\ \ref{fig:MoTe2_SnS2_TR} (d). The band-to-band tunneling is possible in the bilayer for the inverted energy window of -3.94 and -3.86 eV and the transmission peaks show up as seen in Fig.\ \ref{fig:MoTe2_SnS2_TR} (e) and (f). Again notice the good agreements between the MS solutions and full solutions both in the linear and log scales.

\subsection{\label{sec:InAs_trap}InAs nanowire FET with a defect}

\begin{table}
\caption{\label{table:InAs_trap}
The full-Hamiltonian size $N_b$ and the number of modes $n_b$ of the cells of the supercell of Fig.\ \ref{fig:InAs_trap} (a). A total of 28 unphysical branches were cleared during the hetero-structure MST procedure.}
\begin{ruledtabular}
\begin{tabular}{cccccc}
cell & A & JA1 & R & JA2 & A \\
\hline
$N_b$ & 2664 & 2664 & 3996 & 2664 & 2664 \\
$n_b$ & 127 & 154 & 172 & 153 & 127
\end{tabular}
\end{ruledtabular}
\end{table}

\begin{figure*}
\includegraphics[width=\textwidth]{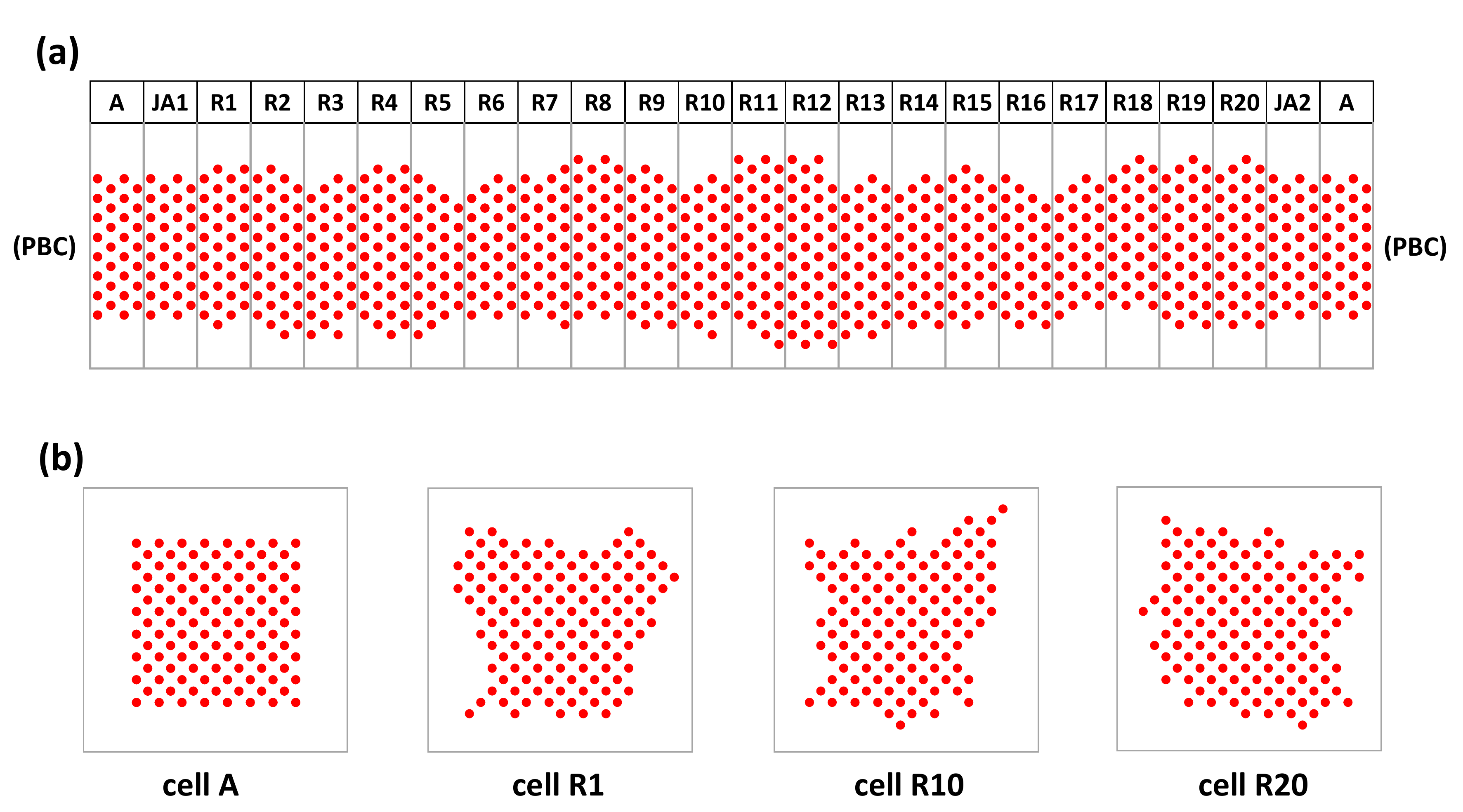}
\caption{\label{fig:Si_SR} Silicon nanowire with rough surfaces: (a) The lateral view of the 24-cell supercell used in the hetero-structure MST. The cells in the hetero region from cell R1 through R20 have rough surfaces. The periodic boundary condition is imposed on the supercell. (b) The cross-sectional view for cells A, R1, R10, and R20.}
\end{figure*}

As an example for the A-R-B type hetero-structure, we simulated InAs nanowire FET with an As-antisite defect (an In atom is replaced by an As atom) as schematically shown in Fig.\ \ref{fig:devices_examples} (c). The supercell to which the hetero-structure MST was applied is shown in Fig. \ref{fig:InAs_trap} (a), where there are 5 cells marked with A, JA1, R, JA2, A cells. The cell A is regarded (and verified to be) {\it unit cell}, the cell R contains the As-antisite defect, and JA1 and JA2 are the junction cells which serve as buffer cells to block the effect of the cell R on the cell A. 

The nanowire structure is oriented in such a way that the transport direction is parallel to [110] direction. The cross-section of the nanowire has six facets with the diameter of 2.4 nm. The facet where the As-antisite is created in the cell R was passivated with Al and O atoms by employing the so-called atom-deposition-layer-like passivation, while the rest facets were passivated with H atoms. The size of the cell A and the junction cells JA1 and JA2 is twice the primitive unit cell of homogeneous structure. There are 184 atoms in each of cells A, JA1, and JA2; 64 In atoms, 64 As atoms, 6 Al atoms, 4 O atoms, and 46 H atoms. The cell R which contains the defect is larger than other cells, containing 95, 97, 9, 6, and 69 atoms of In, As, Al, O, and H, respectively (total 276 atoms). The cells are larger than primitive unit cells because the interaction range between atoms is greater than 5th neighbor and the supercell Hamiltonian is required to have block tri-diagonal form in this work. 

The supercell was relaxed by DFT using the SIESTA tool in the same way as the one in Section \ref{sec:GaSb_InAs}. With the relaxed supercell, we carried out the hetero-structure MS procedures. The resulting number of modes are shown in Table \ref{table:InAs_trap} for the energy window of VBE-0.7 to CBE+0.7. The MS Hamiltonian was reduced to around 5 \% of its original size. 

With the cells extracted from the relaxed supercell, we constructed the structure for transport calculation as shown in Fig.\ \ref{fig:InAs_trap} (c). For the potential profile shown in Fig.\ \ref{fig:InAs_trap_TR} (a), where there is an energy window of 0.3 eV for band-to-band tunneling, we calculated the transmission and LDOS at each cell as shown in Fig.\ \ref{fig:InAs_trap_TR} (b) and (c) and compared them with the full solutions. In Fig.\ \ref{fig:InAs_trap_TR} (b), the energy-resolved transmission shows a peak at -6.56 eV which coincides with the energy of the trap state due to the As-antisite defect. LDOS at each cell also show a peak at the trap energy. The hetero-structure MS and full calculations match excellently for both the transmission and LDOS. In a closer look, the trap state is seen to be spread by about 0.005 eV in energy, which is also faithfully reproduced by the MS Hamiltonian.

\begin{figure}
\includegraphics[width=\columnwidth]{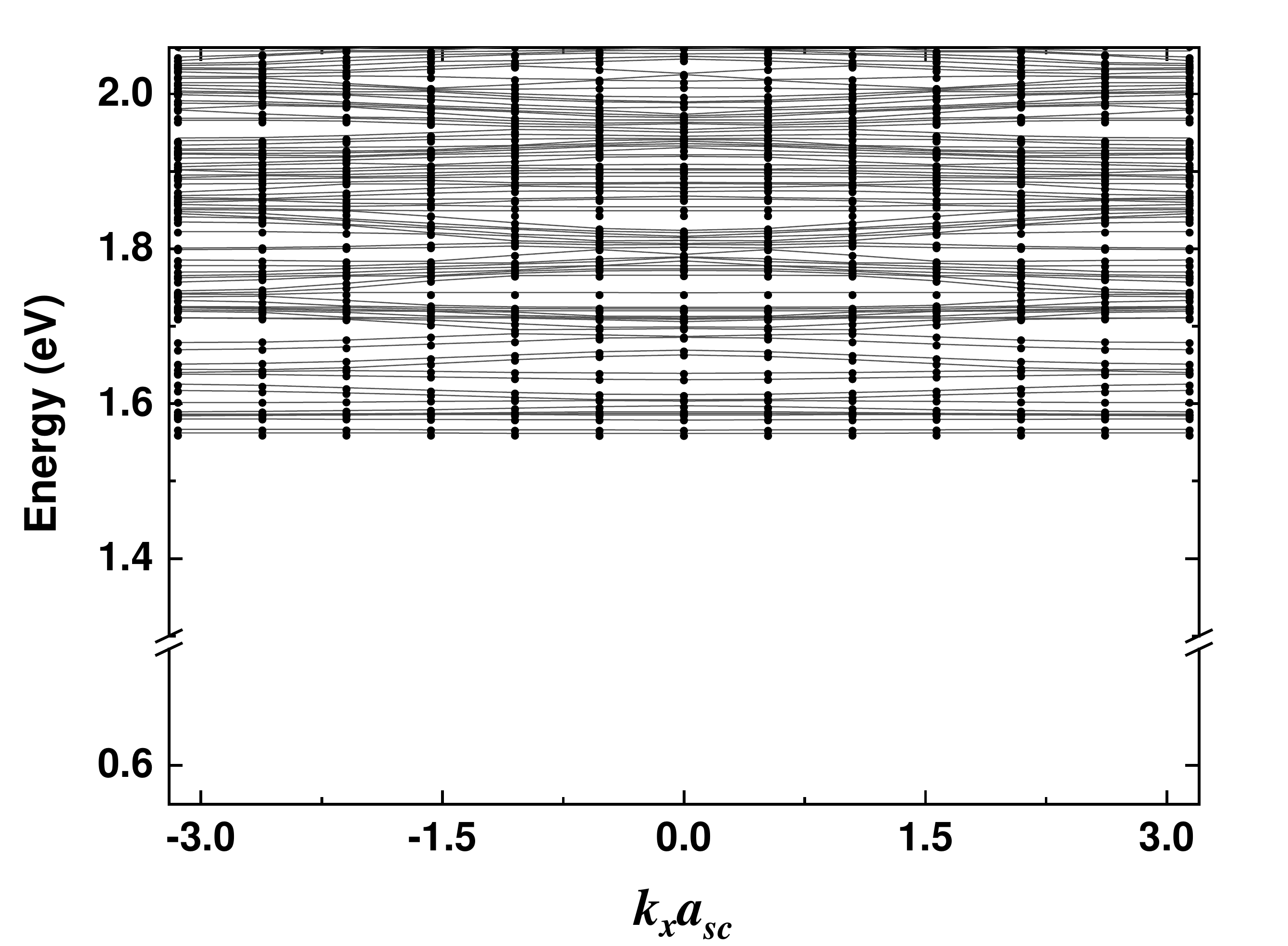}
\caption{\label{fig:Si_SR_band} The band structure of the supercell of Fig.\ \ref{fig:Si_SR} calculated by using the MS Hamiltonian (thin lines) and by using the full Hamiltonian (solid dots) in the energy window of interest from 0.56 to 2.06 eV which corresponds to CBE-1.0 and CBE+0.5, respectively.}
\end{figure}

\begin{figure}
\includegraphics[width=\columnwidth]{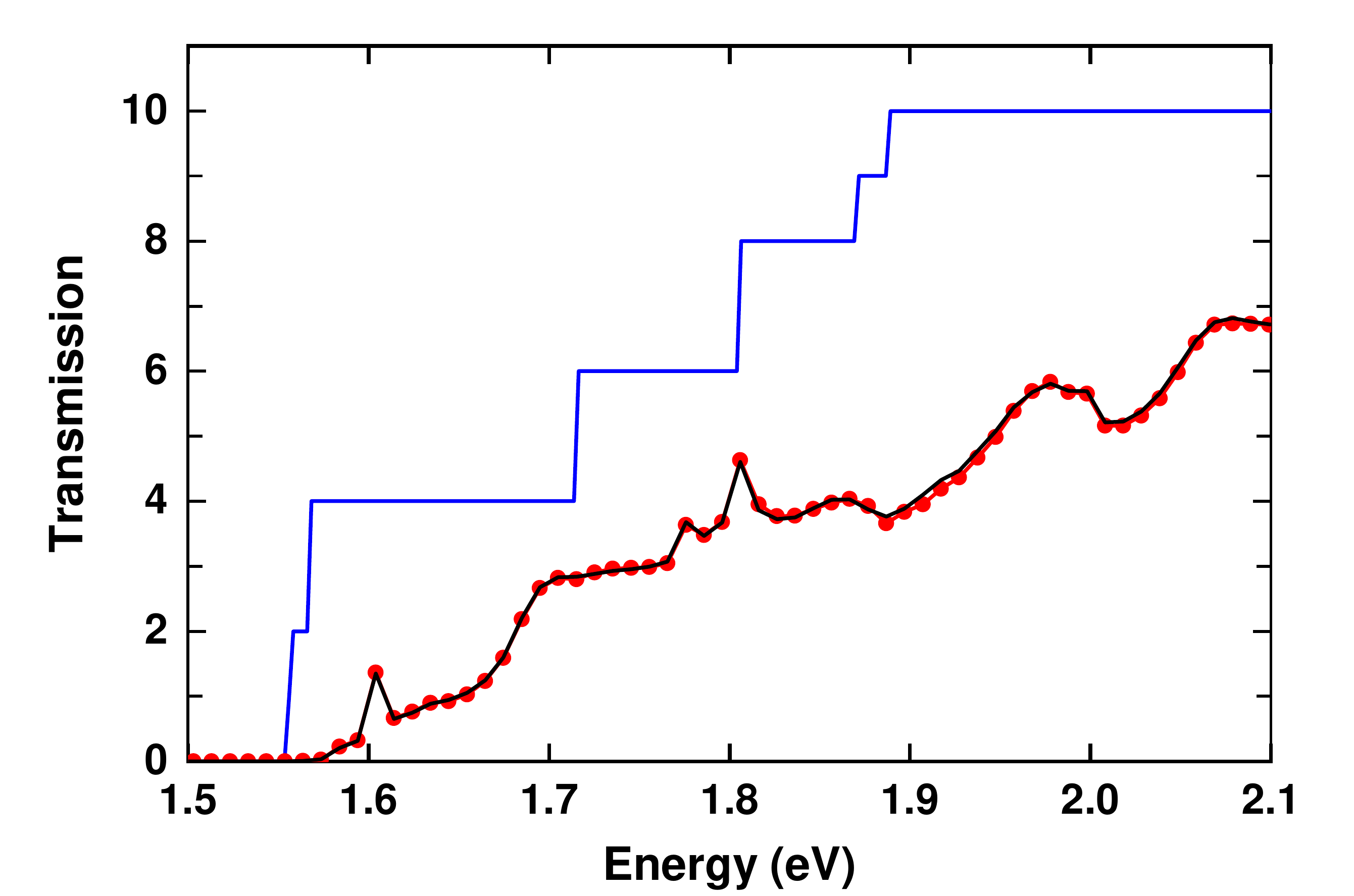}
\caption{\label{fig:Si_SR_TR} The transmission of the hetero-structure of Fig.\ \ref{fig:Si_SR} calculated by using the MS Hamiltonan (red dots) and by using the full Hamiltonian (black line). The transmission of the homogeneous structure without the surface roughness in the channel region is shown as a reference (blue line). }
\end{figure}

\subsection{\label{sec:Si_SR}Silicon nanowire with surface roughness}

As an extreme example of the A-R-B type hetero-structure simulation, we simulated Si nanowire with rough surface in the channel region as shown in Fig.\ \ref{fig:devices_examples} (d). 
Fig.\ \ref{fig:Si_SR} (a) shows the supercell to which the hetero-structure MST is applied. As we used the empirical $sp^3s^*d^5$ TB model\cite{Boykin2004} with nearest neighbor interaction for this system, the cell A at the leftmost and rightmost is the same as the primitive unit cell of homogeneous Si nanowire of 2 nm $\times$ 2 nm cross-section and only one junction cell is needed at each of the left and right junctions of the supercell. In the 11 nm long channel region consisting of 20 cells, the surface roughness is realized at four Si/oxide interfaces using the exponentially decaying auto-correlation function expressed as\cite{Goodnick1985,Jung2013}
\begin{equation}
    C(\vec{r}) = \Delta_m^2 \exp(-\sqrt{2} r/L_m)
\end{equation}
where $\Delta_m$ is the root mean square fluctuation of the roughness, $L_m$ is the correlation length, and $r =|\vec{r}|$ is the distance between two points on the interface. All the danging bonds are implicitly passivated with hydrogen. Fig.\ \ref{fig:Si_SR} (a) shows an instance of generated rough surface. We intentionally used $\Delta_m$ of 0.3 nm and $L_m$ of 1 nm to generate severely rough surface as evidenced by the cross-sectional view of some selected cells in Fig.\ \ref{fig:Si_SR} (b).

The hetero-structure MST was applied to the 24-cell supercell. The MS Hamiltonian was constructed to reproduce the conduction band only, with the energy window of CBE-1.0 and CBE+0.5. The resulting number of modes are shown in Table \ref{table:Si_SR}. The average number of modes is around 48, which is less than 5 \% of the average size of the full Hamiltonian of 1130. Fig.\ \ref{fig:Si_SR_band} shows the band structure of the supercell after the hetero-structure MS procedure is completed with all the unphysical states cleared in the energy window of interest. The resulting bands match excellently with the bands calculated by using the full Hamiltonian. This supports the ansatz mentioned in Section \ref{sec:initial_basis} that the initial basis matrix of the form in Eq. (\ref{eq:U0_construction}) contains all the physical states.

The supercell structure was directly used for transmission calculation. In other words, the same cells were used for the transmission calculation of the device with 11 nm long channel region with surface roughness, but the difference lies in that, for the supercell, the periodic boundary is assumed at both the end cells, but for the transmission calculation, the end cells are assumed to be extended semi-infinitely, which is taken care of by the contact self energies of the NEGF formalism.

Fig.\ \ref{fig:Si_SR_TR} shows the transmission calculated by using the MS Hamiltonian and the one by using the full Hamiltonian. Excellent match between the two is observed. Due to the severely rough surface, the transmission is about half of the step-like transmission of homogeneous Si nanowire with no surface roughness.

\begin{table}
\caption{\label{table:Si_SR}
The full-Hamiltonian size $N_b$ and the number of modes $n_b$ of the cells of the supercell of Fig.\ \ref{fig:Si_SR} (a). A total of 49 unphysical branches were cleared during the hetero-structure MST procedure.}
\begin{ruledtabular}
\begin{tabular}{ccccccccccc}
Cell & A & JA1 & R1 & R2 & R3 & R4 & R5 & R6 & R7 &R8\\
\hline
$N_b$ & 1130 & 1130 & 1260 & 1240 & 1270 & 1310 & 1130& 1080 & 1100 & 1100 \\
$n_b$ & 51 & 51 & 53 & 52 & 51 & 50 & 48 & 46 & 44 & 45\\[0.1cm]
\hline
Cell & R9 & R10 & R11 & R12 & R13 & R14 & R15 & R16 & R17 & R18\\
\hline
$N_b$ & 1300 & 1190 & 1350 & 1520 & 1200 & 1210 & 1160 & 1190 & 1030 & 1150 \\
$n_b$ & 47 & 48 & 48 & 49 & 48 & 46 & 45 & 42 & 43 & 45\\[0.1cm]
\hline
Cell & R19 & R20 & JA2 & A \\
\hline
$N_b$ & 1320 & 1320 & 1130 & 1130 \\
$n_b$ & 49 & 49 & 51 & 51\\
\end{tabular}
\end{ruledtabular}
\end{table}

\section{\label{sec:conclusion}Conclusions\protect}

The ballistic transport in the devices has been assumed for the demonstrations in this paper. However, as an effective Hamiltonian is constructed which can be regarded virtually equivalent to the original Hamiltonian within a preset energy window of interest, it can be readily used to describe the dissipative transport as well. The hetero-structure MST applied to the phonon-assisted tunneling problem in InAs nanowire FET will be published elsewhere.

 It is true that the unphysical-state-clearing procedure for the supercell consisting of many cells takes considerable CPU time. Using one 32-core CPU, it took only a few minutes to clear all the unphysical states for the GaSb/InAs hetero-structure example, but it took about two hours for the rough-surface Si nanowire example where there are 24 cells in the supercell. Although it is tiny compared to the time it takes for the transport calculation with the original full-sized Hamiltonian, there is a room for improvement in terms of the minimization method, alternative trial functions, and parallelization to name a few. Still, there is a huge advantage of working in the MS space in terms of memory and CPU usage in the transport calculation.

In summary, a MS method for hetero-structure device simulations has been developed and its efficiency and accuracy have been demonstrated in this work. This method is promising in that a wide range of hetero-structures can be efficiently simulated with the freedom to employ any type of Hamiltonian. For instance, the old yet important problem of Schottky junctions at the atomistic scale, nanoscale metal interconnects, various novel memory devices consisting of ferroelectrics or chalcogenides or oxides with conducting ions sandwiched by metals can be tackled by the method in this work.

\begin{acknowledgments}

The Hamiltonians used in this work are mainly from a data archive in the School of Electrical Engineering, KAIST. M.\ Shin acknowledges Yongsoo Ahn, Yoon Hee Chang, Yucheol Cho, Seong Hyeok Jeon, and Yeongjun Lim for their contributions to the archive. M.\ Shin specially thanks Yeongjun Lim for his drawing of the atom images featured in this paper. This work was supported by Samsung Research Funding and the Incubation Center of Samsung Electronics under project number SRFC-TA1703-10.

\end{acknowledgments}

The authors have no conflicts to disclose.

\section*{data availability}
The data that supports the findings of this study are available within the article.

\bibliography{HeteroRBT}

\end{document}